\documentclass[10pt,letterpaper]{article}
\usepackage[T1]{fontenc}
\usepackage{array}
\usepackage{amsmath}
\usepackage{amssymb}
\usepackage{graphicx}
\usepackage{setspace}

\makeatletter

\providecommand{\tabularnewline}{\\}

\usepackage{opex3}
\usepackage{cite}

 \let\maketitle=\relax
\makeatother

\begin{document}

\title{Measurements of the dependence of the photon-number distribution
on the number of modes in parametric down-conversion}

\author{L. Dovrat, M. Bakstein, D. Istrati, A. Shaham, and H. S. Eisenberg*}

\maketitle
\address{Racah Institute of Physics, Hebrew University of Jerusalem, Israel}

\email{*hagai.eisenberg@huji.ac.il}
\begin{abstract}
Optical parametric down-conversion (PDC) is a central tool in quantum
optics experiments. The number of collected down-converted modes greatly
affects the quality of the produced photon state. We use Silicon Photomultiplier
(SiPM) number-resolving detectors in order to observe the photon-number
distribution of a PDC source, and show its dependence on the number
of collected modes. Additionally, we show how the stimulated emission
of photons and the partition of photons into several modes determine
the overall photon number. We present a novel analytical model for
the optical crosstalk effect in SiPM detectors, and use it to analyze
the results.
\end{abstract}

\section{Introduction}

For more than two decades, the optical process of parametric down-conversion
(PDC)\,\cite{MandelWolf_OpticalCoherence} has been the primary photon
source in quantum optics experiments\,\cite{Kwiat1995}. This nonlinear
process converts single photons into pairs of signal and idler photons,
which possess quantum correlations between their various degrees of
freedom. The statistical nature of this process is a subject of great
interest, as the photon-number statistics of the process provide a
direct indication of its non-classility\,\cite{Waks2004,Waks2006,Ivanova2006,Lee1990_2modes}.
Additionally, the photon statistics have practical implications on
state preparation schemes, where the quality of the produced state
depends on the number of photons generated simultaneously within the
PDC process\,\cite{Eisenberg2004}.

Recent developments in photon-number resolving detectors\,\cite{Lincoln2000_VLPC,Achilles2004,Rosenberg2005,Kardynal2008,DaulerBerggren2009}
enable the implementation of a wide range of applications in quantum
optics, including state preparation schemes\,\cite{KokDowling2002,GaoDowling2010,Eisenberg2005},
quantum state filtering\,\cite{Okamoto2009} and quantum gates\,\cite{KLM2001,Pittman2002}.
Number discriminating capabilities have also been shown to aid in
probing eavesdropping attacks on quantum cryptography protocols\,\cite{Durkin2002}.
Furthermore, one of the most prominent advantages of number-resolving
detectors is the ability to directly measure the photon-number statistics
of an optical state.

The photon-number distribution of a PDC source can range between a
thermal distribution, when a single down-converted mode is collected,
and a Poisson distribution when an infinitely large number of spatial
and spectral modes (multimode) is collected\,\cite{Mandel1959}.
The photon statistics of PDC sources has been directly measured in
several works\,\cite{Paleari2004,Waks2004,Waks2006,Avenhaus2008},
all of which report a Poisson distribution. However, the majority
of quantum optics schemes take great care to ascertain that only a
single mode is collected. Using an indirect homodyne detection method,
a thermal distribution from a parametric amplifier was demonstrated\,\cite{Vasilev1998}.
Transition between thermal to Poisson distributions was indirectly
observed through the $g^{(2)}$ parameter\,\cite{Eckstein2011} and
through high-order PDC events\,\cite{Wasilewski_Banaszek2008}. In
this work, we use a photon-number resolving detector in order to directly
measure the photon statistics of the signal and idler photons collected
from a controlled number of modes, allowing us to observe the range
of distributions between the single-mode and the multimode extremes.

The photon-number resolving detector used in this work is the Silicon
Photomultiplier (SiPM)~\cite{Bondarenko1998}. The SiPM detector
is composed of multiple avalanche-photo-diodes (APD) operating in
Geiger-mode, combined on a single substrate. When a photon impinges
on an APD element, an electric discharge can be generated. The photons
impinging on the detector are distributed across its many elements.
The output signals generated by all APDs are combined to form a single
output pulse whose intensity is proportional to the number of triggered
elements. In comparison to other technologies, the SiPM detector offers
a better photon-number resolution (up to 20 photons, compared to 10
or less), easy integration in optical setups and operation at room
temperature. On the other hand, its photon detection efficiency is
2--6 times lower, depending on the working wavelength.

The photon statistics measured by SiPM detectors may deviate from
that of the original state. The three main reasons for these deviations
are: 1)\ the non-perfect overall photon detection efficiency (loss),
which is determined by the detection and coupling efficiencies, 2)\ false
detections caused by thermal excitations (dark counts), and 3)\ optical
crosstalk, in which secondary photons, created by carrier relaxation
in one APD element, are detected by a neighboring element, falsely
increasing the photon count\,\cite{Buzhan2006}. Another deviation
is caused by events when two or more photons impinge on the same element
and are detected as one. However, we have shown using a numerical
model that this effect becomes significant only when the number of
detected photons is higher than 20\% of the total number of elements\,\cite{DovratSiPM2011}.
The descriptions of loss and dark counts are quite straightforward.
However, several currently available models for the crosstalk effect
are either computationally difficult\,\cite{DovratSiPM2011} or only
applicable under limiting conditions\,\cite{Afek2009,Akiba2009,Eraerds2007}.
We introduce here an analytical crosstalk model which can be applied
for the entire range of experimental settings.

This paper is organized as follows: In Sec.~\ref{sec:Data-Analysis}
we present a model for the distortion effects in the SiPM detector,
that includes a new approach to the crosstalk effect. This model is
used for the interpretation of experimental data and the reconstruction
of the original photon-number statistics. Sec.~\ref{sec:Experimental}
describes the experimental setup for the generation of PDC states
and their detection using an SiPM photon-number resolving detector.
In Sec.~\ref{sec:Results-&-Discussion} we present measurements of
photon-number distributions of parametric down-conversion and their
dependence on the number of collected modes. Conclusions are presented
in Sec.~\ref{sec:Conclusions}.

\section{The SiPM detection model \label{sec:Data-Analysis}}

In order to interpret the photon-number statistics measured using
SiPM detectors, the effects of loss, dark counts, and optical crosstalk
must be properly modeled. If the measured photon-number distribution
is represented by a vector of probabilities $\vec{p}_{m}$, the original
probability distribution $\vec{p}_{o}$, is related to the measured
distribution through the relation
\begin{equation}
\vec{p}_{m}=\textrm{M}\cdot\vec{p}_{o}\,.\label{eq:measured-orig relation}
\end{equation}
Each component of the distortion matrix $\textrm{M}$ is the conditional
probability $\textrm{M}_{nm}\equiv\textrm{M}(n|m)$ of measuring $n$
photons given that $m$ original photons arrived at the detector.
Using Eq.~\ref{eq:measured-orig relation}, the original distribution
can be reconstructed, in principle, by multiplying the measured distribution
with the inverse of the distortion matrix. This method, however, is
limited to relatively large probabilities and may produce non-physical
results otherwise\,\cite{LeeDowling2004}. Another approach, which
we use here, is to apply a fitting algorithm which finds the original
distribution based on a goodness of fit test.

The matrix $\textrm{M}$ can be written as a product of three matrices,
$\textrm{M}=\textrm{M}_{ct}\cdot\textrm{M}_{dk}\cdot\textrm{M}_{loss}$,
which represent the loss, dark counts (dk) and crosstalk (\textit{ct})
effects. The individual matrices are constructed as follows:

\subsection*{Loss}

The probability for loss is described by the relation\,\cite{LeeDowling2004}
\begin{equation}
\textrm{M}_{loss}(n|m)=\begin{cases}
\left(\begin{array}{c}
m\\
n
\end{array}\right)\eta^{n}\left(1-\eta\right)^{m-n} & n\leq m\\
0 & n>m\,,
\end{cases}\label{eq:Loss Matrix}
\end{equation}
where $\eta$ is the overall detection efficiency.

\subsection*{Dark counts}

The probability of dark counts is assumed to follow a Poisson distribution\,\cite{LeeDowling2004}
of the form
\begin{equation}
\textrm{M}_{dk}(n|m)=\begin{cases}
0 & n<m\\
\frac{\lambda_{dk}^{n-m}\exp(-\lambda_{dk})}{(n-m)!} & n\geq m\,,
\end{cases}\label{eq:DK matrix}
\end{equation}
where $\lambda_{dk}$ is the average number of dark counts.

\subsection*{Crosstalk}

Optical crosstalk occurs when a secondary photon generated during
an electronic discharge in one APD element is detected in a neighboring
APD element, thereby creating a secondary avalanche. Photons produced
in the secondary avalanches can then continue to trigger further avalanches
in additional elements (crosstalk-generated-crosstalk). The crosstalk
probabilities in our model are calculated by dividing the crosstalk
events into stages and recursively counting the number of possible
crosstalk events in each stage. The recursion formula for the crosstalk
matrix element is:
\begin{equation}
\textrm{M}_{ct}(n|m)=\begin{cases}
\sum\limits _{n_{ct}=0}^{min(m,n-m)}\left(\begin{array}{c}
m\\
n_{ct}
\end{array}\right)\cdot\epsilon^{n_{ct}}(1-\epsilon)^{m-n_{ct}}\cdot\textrm{M}_{ct}(n-m|n_{ct}) & n\geqslant m>0\\
1 & n=m=0\\
0 & \textrm{otherwise}\,,
\end{cases}\label{eq:CT matrix}
\end{equation}
where $\epsilon$ is defined as the overall probability for a crosstalk-avalanche
to be generated in any of the four neighboring elements. Each stage
has $m$ triggered elements which can trigger up to $m$ new crosstalk
events. The probability of generating $n_{ct}$ new crosstalk events
is composed of the probability that $n_{ct}$ elements out of the
$m$ possible will generate crosstalk, while the remaining $m-n_{ct}$
elements will not. This probability also includes the combinatorial
factor of choosing the $n_{ct}$ crosstalk-generating elements out
of the possible $m$. The $n_{ct}$ crosstalk-triggered elements continue
to generate additional crosstalk events in a recursive way, until
all of the $n-m$ crosstalk elements are triggered.

This novel analytical crosstalk model does not limit the number of
crosstalk events and also accounts for crosstalk-generated-crosstalk.
The model was developed since other available analytical models~\cite{Afek2009,Akiba2009,Eraerds2007}
were inconsistent with our experimental results due to limitations
on the crosstalk probability values and the number of crosstalk events.
Our model does not account for the geometrical arrangement of the
APD elements in the detector, nor its finite size. However, we show
below that for values within the experimental range, this analytical
model agrees with a numerical model that has no such limitations\,\cite{DovratSiPM2011}.
Additionally, this model is less computationally demanding than the
numerical model. Thus, it is a useful tool in the interpretation of
experimental data from photon-number resolving detectors, where crosstalk
between the detection elements plays a crucial role.

\section{The experimental setup\label{sec:Experimental}}

\begin{figure}
\noindent \begin{centering}
\includegraphics[width=0.7\columnwidth]{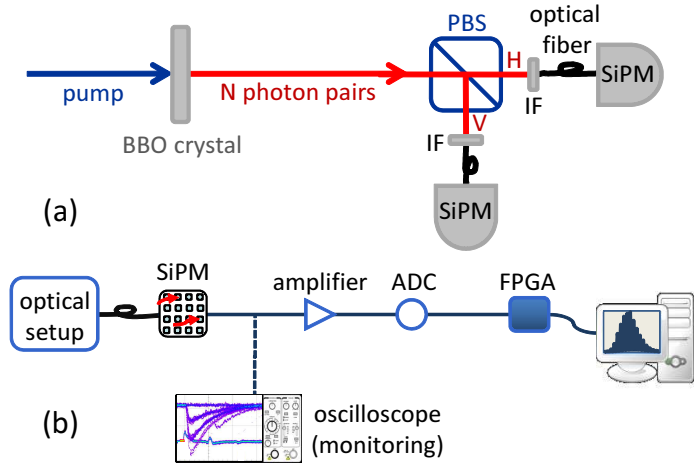}
\par\end{centering}

\noindent \centering{}\caption{(a) The experimental setup. A type-II collinear BBO crystal is pumped
by 390~nm amplified pulses at a repetition rate of 250\,kHz. The
signal and idler photons are split at a Polarizing Beam Splitter (PBS)
according to their polarization and detected by two SiPM detectors.
The number of spatial and spectral modes which are collected is varied
using interference filters (IF) with different bandwidths and different
optical fibers. (b) The detection configuration. An SiPM detector produces
a signal whose intensity is proportional to the number of impinging
photons. This signal is amplified, digitized using analog-to-digital
converters (ADC) and analyzed using FPGA electronics in real-time.
The data is continuously transmitted to a computer, which displays
the photon-number distribution.\label{fig:setup}}
\end{figure}

We have measured the photon-number distributions of the polarization
modes of a stimulated type-II collinear PDC process. The experimental
setup is shown in Fig.~\ref{fig:setup}(a). Pulses of 150\,fs at
a repetition rate of 250\,kHz are amplified, frequency doubled to
390\,nm and used to pump a 2\,mm long type-II collinear $\beta-\textrm{BaB\ensuremath{_{2}}O\ensuremath{_{4}}}$
(BBO) nonlinear crystal. The degenerate wavelength signal and idler
photons at 780\,nm, which are created with orthogonal polarizations
(horizontal (H) and vertical (V)), are split at a polarizing beam
splitter (PBS) and detected separately using two photon-number resolving
detectors. This work focuses on measuring the photon-number distribution
of a single polarization mode. Thus, the detection of both polarization
modes is only used to evaluate the heralded detection efficiency of
the SiPM detector. The results presented for a single polarization
mode also apply to the photon pair statistics, as a collinear process
produces the same statistics for pairs and for individual photons.
Before the photons are coupled to the detectors, they are spectrally
and spatially filtered by bandpass filters and by optical fibers,
respectively. The number of spatial modes which are collected can
be changed by using optical fibers with different core diameters and
numerical apertures\,\cite{YarivCommunications}, and the number
of collected spectral modes can be changed by using filters of different
bandwidths.

The detection configuration is shown in Fig.~\ref{fig:setup}(b).
The optical signal from the photon source is coupled to an SiPM detector
with 100 elements (\emph{Hamamatsu Photonics}, S10362-11-100U) using
optical fibers. The output electrical signal from the detector is
amplified using low-noise amplifiers, digitized and then analyzed
by FPGA electronics, where the number of detected photons is extracted
in real time. This data is continuously transmitted to a computer,
which presents the statistics. In order to minimize the effects of
dark counts and after-pulsing\,\cite{Buzhan2006}, the 1~ns long
sampling time of the analog electrical signal is synchronized with
the arrival time of the photons. The dark counts are further reduced
by moderately cooling the detectors using a thermoelectric cooler
to $\approx-10^{\circ}\textrm{C}$, and the bias voltage level applied
to the SiPM is modified accordingly.

\section{Results and discussion\label{sec:Results-&-Discussion}}

\begin{figure}
\noindent \begin{centering}
\includegraphics[width=0.7\columnwidth]{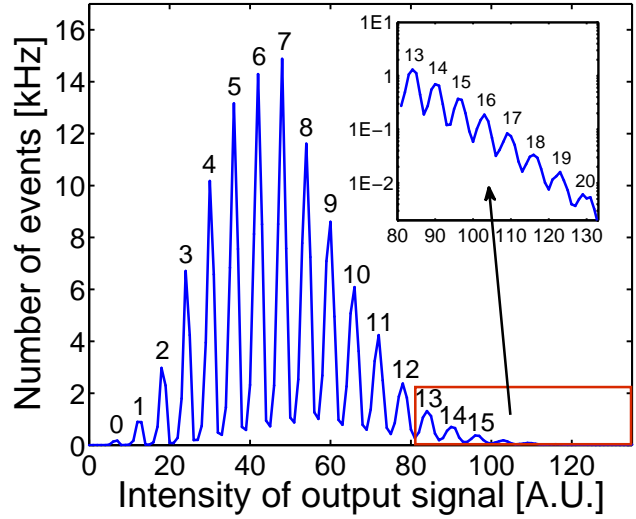}
\par\end{centering}

\caption{A histogram of the electrical output signal level for a coherent input
state. The data was accumulated over a period of 30 seconds. Good
peak separation is maintained up to 20 photons. \label{fig:Histogram}}
\end{figure}

Figure~\ref{fig:Histogram} shows an example of a histogram of pulse
intensities produced by a coherent state. The probability of detecting
$n$ photons is proportional to the area of the $n^{th}$ Gaussian.
We obtain high photon-number discrimination for as high as 20 photons.
The photon-number resolution error, which results from the overlapping
between neighboring Gaussians, is smaller than 1\% below 12 photons
and approaches 12\% for 20 photons. The number of resolvable photons
is limited due to the decrease of the signal to noise ratio as the
number of photons is increased.

\begin{figure}[t]
\noindent \begin{centering}
\includegraphics[width=0.7\columnwidth]{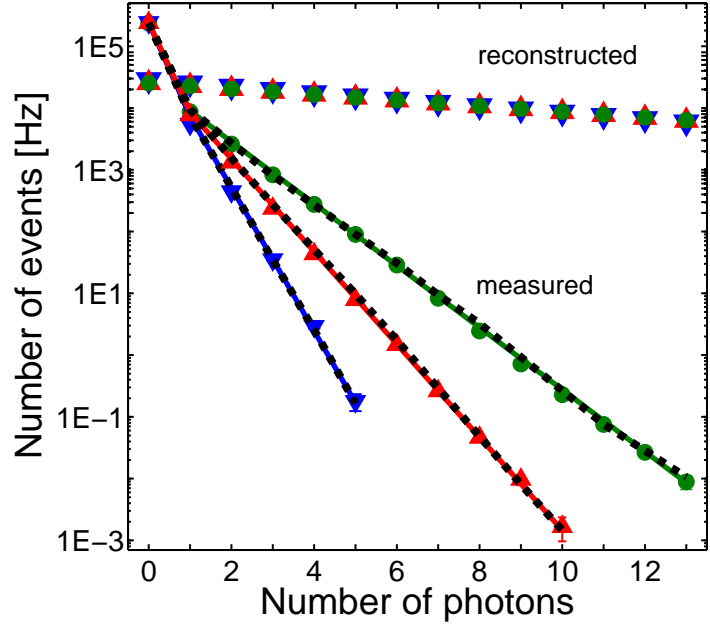}
\par\end{centering}

\caption{Measurements of a thermal state conducted on a polarization mode of
a type-II collinear PDC source, spatially and spectrally filtered
using a single-mode fiber and a 3\,nm bandpass filter. The experimental
data was fitted to the obtained statistics using the model presented
in this work (solid lines). The reconstruction of the original photon
statistics from the measured data is also shown. The data was obtained
at a temperature of $-10^{\circ}\textrm{C}$. The bias voltage values
and the fit parameters are: $V_{bias}=68.3\,\textrm{V}$, $\eta=6\cdot10^{-3}\pm1\cdot10^{-3}$,
$\lambda_{dk}=2.4\cdot10^{-3}\pm6\cdot10^{-4}$, $\overline{n}=8.9\pm0.5$,
$\epsilon=0.280\pm0.005$ (green circles), $V_{bias}=67.7\,\textrm{V}$,$\eta=4.9\cdot10^{-3}\pm3\cdot10^{-4}$,
$\lambda_{dk}=1.0\cdot10^{-3}\pm3\cdot10^{-4}$, $\overline{n}=8.8\pm0.5$,
$\epsilon=0.140\pm0.006$ (red upward triangles), and $V_{bias}=67.4\,\textrm{V}$,
$\eta=4.2\cdot10^{-3}\pm5\cdot10^{-4}$, $\lambda_{dk}=2.5\cdot10^{-4}\pm1\cdot10^{-4}$,
$\overline{n}=7.8\pm2.7$, $\epsilon=0.040\pm0.006$ (blue downward triangles).
The corresponding crosstalk values obtained with the numerical model
of Ref.\,\cite{DovratSiPM2011} are $\epsilon_{nn}=0.078\pm0.001$,
$\epsilon_{nn}=0.038\pm0.001$, and $\epsilon_{nn}=0.010\pm0.001$,
respectively. Fits using the numerical model are presented by black
dashed lines.\label{fig:model fits}}
\end{figure}

We demonstrate the use of our model for reconstructing the original
photon statistics with the photon-number distribution measurements
of a thermal state. The thermal state was generated by collecting
a single polarization mode of PDC using the setup shown in Fig.~\ref{fig:setup}.
The original thermal distribution can be written as
\begin{eqnarray}
p_{th}(n) & = & \frac{1}{\left(1+\overline{n}\right)\left(1+\frac{1}{\overline{n}}\right)^{n}}\,,\label{eq:pth_thermal}
\end{eqnarray}
where $\overline{n}$ is the average number of photons. The measured distribution
exhibits a lower value of $\overline{n}$ due to losses and deviates from
the power-law dependence of Eq.~\ref{eq:pth_thermal} due to optical
crosstalk and dark counts. We measured the same thermal state with
three different sets of values for the amount of loss, dark counts
and crosstalk, obtained by operating the SiPM detector with different
bias voltage values\,\cite{Buzhan2006}. As a result, we obtained
three different measured distributions, although the original photon-number
distribution was the same. The measurements are shown in Fig.~\ref{fig:model fits}.
The main distortions in the distributions are the low values of $\overline{n}$
due to losses and the deviation from a straight line in the semi-log
plot when the number of measured photons exceeds 1. The latter is
the effect of the crosstalk process, which falsely increases the probability
of measuring high numbers of photons. The crosstalk probabilities
of these measurements range from the highest probability value (maximum
bias voltage), which creates the highest deviation, to the minimal
probability (minimal bias voltage) which creates the lowest deviation.

The original distribution is reconstructed by fitting the measured
data $\vec{p}_{m}$ to the function $\vec{p}_{m}=\textrm{M}_{ct}\textrm{M}_{dk}\textrm{M}_{loss}\vec{p_{th}}$,
where $\vec{p}_{th}$ is the thermal distribution of Eq.~\ref{eq:pth_thermal}
and $\overline{n}$ is the only free parameter. The overall detection efficiency
$\eta$, the average number of dark counts $\lambda_{dk}$, and the
crosstalk probability $\epsilon$, that are used for the construction
of the distortion matrices, are measured separately. The value of
$\eta$ is defined as the heralded efficiency between the two down-converted
polarization modes and is determined by the ratio between two-photon
coincidence and single-photon counts in the limit $\overline{n}\ll1$.
The dark counts and crosstalk parameters are obtained through a separate
measurement of the Poissonian dark count statistics which are fit
to the function $\vec{p}_{m}=\textrm{M}_{ct}\textrm{M}_{dk}\vec{\mathsf{1}}$,
where $\vec{\mathsf{1}}$ is a vector of zeros with 1 at the first
position, and $\lambda_{dk}$ and $\epsilon$ are the free fit parameters.

All three measurements in Fig.~\ref{fig:model fits} were reconstructed
to the same original state, within the margins of error, despite the
different measured distributions. This result indicates the consistency
of our evaluation of the dark counts, loss and crosstalk parameters.
We compare these results to those obtained using the numerical crosstalk
model introduced in Ref.\,\cite{DovratSiPM2011}. The numerical model
is based on simulations of the detection process, and takes into account
the geometrical configuration of the detector, the finite number of
detection elements, and possible attempts to trigger neighboring cells
which have already been triggered. The model is characterized through
a parameter $\epsilon_{nn}$, defined as the probability of generating
crosstalk in one particular neighbor among the four nearest neighbors,
rather than the overall probability of generating crosstalk among
the nearest neighbors, which we define here as $\epsilon$. The values
of $\epsilon_{nn}$ obtained from the fit correspond to the values
of $\epsilon$ through the relation $\epsilon=1-\left(1-\epsilon_{nn}\right)^{4}$\,\cite{DovratSiPM2011}.
The agreement between the two models shows that although the analytical
model does not account for the geometry of the detector, it provides
a good description of the crosstalk process for all crosstalk values
in the experimental range.

We now consider the photon-number distributions of down-converted
photons as a function of the number of collected modes. If $n$ photons
are distributed among $s$ modes, such that $n_{1}+n_{2}+\dots n_{s}=n$,
then the overall probability of measuring $n$ photons is given by\,\cite{Mandel1959}:
\begin{equation}
p_{s}(n)=\left(\begin{array}{c}
s+n-1\\
s-1
\end{array}\right)\cdot\Pi_{i=1}^{s}p(n_{i})\,,\label{eq:ps(n)}
\end{equation}
where the product $\Pi_{i=1}^{s}p(n_{i})$ is the joint probability
of measuring $n_{i}$ photons in each mode $i$, and the combinatorial
factor amounts to the number of possible arrangements of $n$ indistinguishable
photons into $s$ modes. If the photons are distributed evenly among
the modes, the average number of photons in each mode is uniform and
equals $\overline{n_{i}}=\overline{n}/s.$ The photon-number distribution within
each of the modes is thermal, $p(n_{i})\equiv p_{th}(n_{i})$, and
by substituting Eq.~\ref{eq:pth_thermal} into Eq.~\ref{eq:ps(n)},
we obtain the following probability distribution
\begin{equation}
p_{s}(n)=\left(\begin{array}{c}
s+n-1\\
s-1
\end{array}\right)\frac{1}{\left(1+\frac{\overline{n}}{s}\right)^{s}\left(1+\frac{s}{\overline{n}}\right)^{n}}\,,\label{eq:ps(n) thermal}
\end{equation}
known as the negative binomial distribution. This distribution converges
to a Poisson distribution when the number of modes approaches infinity.

\begin{figure}[tb]
\noindent \begin{centering}
\includegraphics[width=1\columnwidth]{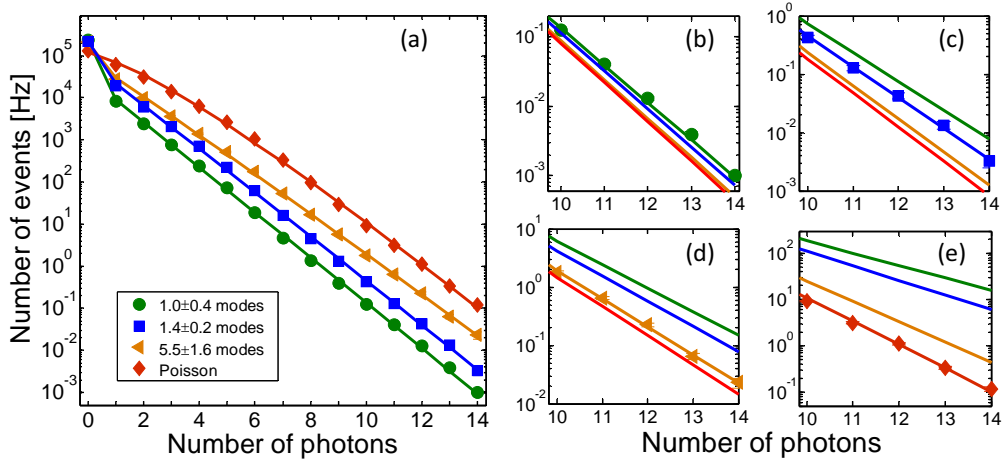}
\par\end{centering}

\caption{(a) Representative photon-number distribution measurements and their
fits. The results are presented for distributions which range between
thermal and Poissonian statistics. Experimental errors are smaller
than their respective symbol size. The average photon number for each
data set is reduced as less modes are collected. (b--e) A close-up
of each of the measured distributions along with distributions calculated
for the same fit parameters while changing the mode-number value between
a single, 1.4 and 5.5 modes, and a Poisson distribution. The measurements
were taken with the following parameters: $\eta=0.013$, $\lambda_{dk}=1.95\cdot10^{-3}$,
$\epsilon=0.26$, 3\,nm bandwidth filter, single mode fiber for 780\,nm,
integration time of 780\,minutes (green circles), $\eta=0.019$,
$\lambda_{dk}=2.1\cdot10^{-3}$, $\epsilon=0.23$, 3\,nm bandwidth
filter, single mode fiber for 1500\,nm, integration time of 100\,minutes
(blue squares), $\eta=0.013$, $\lambda_{dk}=1.69\cdot10^{-3}$, $\epsilon=0.27$,
10\,nm bandwidth filter, single mode fiber for 1500\,nm, integration
time of 100\,minutes (orange triangles), and $\eta=0.020$, $\lambda_{dk}=3\cdot10^{-3}$,
$\epsilon=0.22$, 3\,nm bandwidth filter, multimode fiber for 780\,nm,
integration time of 115\,minutes (red diamonds). \label{fig:mode graph}}
\end{figure}

The dependence of the photon-number distribution on the number of
modes is demonstrated by measuring the photon statistics collected
from a controlled number of modes. Figure~\ref{fig:mode graph} shows
four representative measurements of the photon-number distributions
which range between a single mode (thermal) distribution, and a multimode
(Poisson) distribution. The solid lines are fits to the function $\vec{p}_{m}=\textrm{M}_{ct}\textrm{M}_{dk}\textrm{M}_{loss}\vec{p_{s}}$,
where $\vec{p}_{s}$ is defined in Eq.~\ref{eq:ps(n) thermal} with
$\overline{n}$ and $s$ as the free parameters. We detected up to 14 photons
and obtained good fits to the data. Notice that the difference between
the distributions becomes more marked as the number of photons increases
(see Figs.~\ref{fig:mode graph}(b)--(e)). The difference is almost
undetectable for low photon numbers. In fact, measurements of at least
10 photons are required for our experimental parameter range, in order
to properly discriminate between distributions with different mode
numbers.

The number of modes obtained for all combinations of fibers and filters
is shown in table~\ref{tab:mode table}. The non-integer mode numbers
should be interpreted as a weighted number of occupied modes and result
from a non-uniform distribution of photons between the modes. We can
discriminate between distributions containing up to about 10 modes,
as distributions containing over 10 modes cannot be distinguished
from Poisson distributions. Table~\ref{tab:mode table} shows the
proportional increase of the number of spectral modes with the increase
in the bandwidth of the bandpass filters. The width of the phase-matched
spectrum of the down-converted photons is estimated to be $\approx10\,\textrm{nm}$.
Due to spectral distinguishability, this width includes about 3 temporal
modes\,\cite{Mosley2008,Wasilewski_Banaszek2008,Eckstein2011}. The
number of spatial modes measured with our system is also consistent
with the choice of fibers. When the PDC photons, which passed through
a 3\,nm bandpass filter, were coupled into a single mode fiber for
780\,nm, a thermal distribution was produced. Whereas, when the photons
were coupled into a multimode fiber, a Poisson distribution was produced,
even though the same filter was used. The relatively small number
of modes obtained with the single-mode fiber for 1550\,nm can be
attributed to non-uniform mode coupling, which does not equally stimulate
all of the modes in the fiber.

\begin{table}[t]
\noindent \begin{centering}
\caption{The number of modes as obtained from the photon-number statistics.
The table shows the number of modes for interference filters with
different bandwidths and for optical fibers with different mode field
diameters (MFD) and numerical apertures (NA). Measurements in the
first row were taken with a single-mode fiber for 780\,nm, the second
row with a single-mode fiber for 1550\,nm and the third row with
a standard graded-index multimode fiber. Distributions are considered
Poissonian for $s>10$.\protect \\
\label{tab:mode table}}

\par\end{centering}

\begin{onehalfspace}
\noindent \centering{}%
\begin{tabular*}{5in}{@{\extracolsep{\fill}}>{\centering}p{0.6in}>{\centering}p{0.6in}|>{\centering}p{0.7in}>{\centering}p{0.7in}>{\centering}p{0.7in}>{\centering}p{0.7in}}
\hline
\multicolumn{2}{c|}{fiber type} & \multicolumn{4}{c}{filter bandwidth}\tabularnewline
MFD & NA & \noindent \centering{}3\,nm & \noindent \centering{}5\,nm & \noindent \centering{}10\,nm & \noindent \centering{}no filter\tabularnewline
\hline
 $5.0\,\mu\textrm{m}$ & 0.13 & \noindent \centering{}$1.0\pm0.4$ & \noindent \centering{}$\,\,\,\,1.6\pm0.7$ & $2.6\pm1.6$ & $3.3\pm1.7$\tabularnewline
$10.4\,\mu\textrm{m}$ & 0.14 & \noindent \centering{}$1.43\pm0.24$ & \noindent \centering{}$2.13\pm0.7$ & $5.5\pm1.6$ & \noindent \centering{}$5.4\pm3.2$\tabularnewline
$62.5\,\mu\textrm{m}$ & 0.275 & \noindent \centering{}Poissonian & --- & --- & \noindent \centering{}---\tabularnewline
\hline
\end{tabular*}\end{onehalfspace}
\end{table}

The stimulated nature of PDC can also be used to obtain the number
of collected modes\,\cite{Wasilewski_Banaszek2008}. The average
number of down-converted photons in a single mode is determined by
the stimulation parameter $\tau_{i}$, which has a linear dependence
on the pump field, the crystal length and the nonlinear coefficient
of the crystal\,\cite{Kok2000}. Assuming an equal distribution of
the photons between $s$ modes, the mean number of photons $\overline{n}$
is given by
\begin{eqnarray}
\overline{n} & = & s\cdot\overline{n}_{i}\nonumber \\
 & = & s\cdot\sinh^{2}(\tau_{i})\nonumber \\
 & = & s\cdot\sinh^{2}\left(\alpha\sqrt{I}\right),\label{eq:ns}
\end{eqnarray}
where $\overline{n_{i}}$ is the average number of photons in each of the
individual modes, $\alpha$ is the coupling constant between the pump
field and the nonlinear crystal, and $I$ is the pump intensity.

\begin{figure}[t]
\noindent \begin{centering}
\includegraphics[width=0.7\columnwidth]{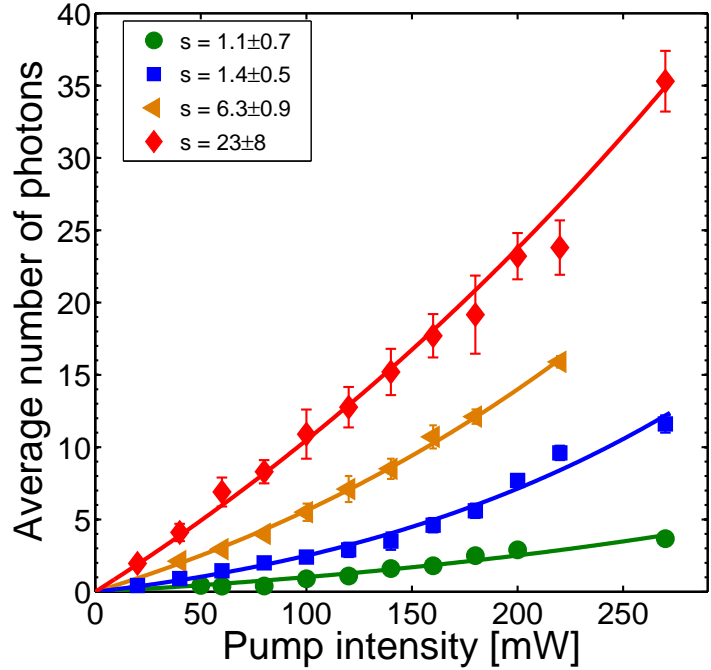}
\par\end{centering}

\caption{The average number of photons collected from multiple modes as a function
of the pump intensity. The solid lines are fits to Eq.~\ref{eq:ns}
with the following parameters: (a) $s=1.1\pm0.7$, $\alpha=0.08\pm0.02$
(green circles) (b) $s=1.4\pm0.5$, $\alpha=0.10\pm0.02$ (blue squares)
(c) $s=6.3\pm0.9$, $\alpha=0.08\pm0.01$ (orange triangles). (d)
$s=23\pm8$, $\alpha=0.06\pm0.01$ (red diamonds).\label{fig:stimulation}}
\end{figure}

In order to show the dependence of $\overline{n}$ on the number of modes,
we measured photon-number distributions as a function of the pump
intensity for the different numbers of collected modes. The values
of $\overline{n}$ were extracted by fitting the different distributions
to Eq.~\ref{eq:ps(n) thermal}. The fit values of $\overline{n}$ as a
function of the total pump power are shown in Fig.~\ref{fig:stimulation}.
All measurements show a nonlinear dependence, indicative of a stimulated
process. The number of modes $s$ obtained from these fits corresponds
well to the number of modes obtained through fits to the photon-number
distributions (see Fig.~\ref{fig:mode graph}). Furthermore, all
fits resulted in similar values of the $\alpha$ parameter, in agreement
with the model of Eq.~\ref{eq:ns}.

\section{Conclusions\label{sec:Conclusions}}

In conclusion, we have built a detection setup, which incorporates
a Silicon Photomultiplier (SiPM) as a photon-number resolving detector.
Our setup enables good photon-number discrimination for as many as
20 photons. The output signal of the SiPM is analyzed in real time
using FPGA electronics. The measured photon-number distribution probabilities
are interpreted by modeling the effects of loss, dark counts and optical
crosstalk, which distort the original photon-number probabilities.
We present a useful analytical crosstalk model that is applicable
for all crosstalk probability values in the experimental range and
includes crosstalk-generated-crosstalk.

We have performed measurements of photon-number distributions of a
single polarization mode of a collinear type-II PDC process. The high
photon-number resolution of our system enables the differentiation
between distributions collected from a different number of spectral
and spatial modes. We present distributions which range from single-mode
(thermal) to multimode (Poissonian) statistics. The results show how
the photon-number distribution is determined by the number of collected
modes. Furthermore, we show the dependence of the average number of
photons on the number of collected modes, and the stimulated nature
of parametric down-conversion.
\end{document}